\documentclass[12pt]{article}
\usepackage{latexsym,graphics,graphicx,amssymb,amsmath, amstext, amsfonts, amsthm, pifont, color, subfig}

\newcommand{\nn}{\nonumber}
\newcommand{\bc}{\begin{center}}
\newcommand{\ec}{\end{center}}
\newcommand{\bfl}{\begin{flushleft}}
\newcommand{\efl}{\end{flushleft}}

\newcommand{\beqa}{\begin{eqnarray}}
\newcommand{\eeqa}{\end{eqnarray}}
\newcommand{\beqan}{\begin{eqnarray*}}
\newcommand{\eeqan}{\end{eqnarray*}}
\newcommand{\beq}{\begin{equation}}
\newcommand{\eeq}{\end{equation}}
\newcommand{\beit}{\begin{itemize}}
\newcommand{\eeit}{\end{itemize}}

\newcommand{\lbr}{\left \{ }
\newcommand{\rbr}{\right \} }

\newcommand{\lp}{\left (}
\newcommand{\rp}{\right )}
\newcommand{\df}{\stackrel{{\rm def}}{=}}

\newcommand{\eqo}{\stackrel{\mbf{.}}{=}}

\newcommand{\real}{{\mathbb{R}}}

\newcommand{\prob}{{\mathbb{P}}}

\newcommand{\mc}{\mathcal}

\newcommand{\mbf}{\mathbf}

\newcommand{\bit}{\begin{itemize}}
\newcommand{\eit}{\end{itemize}}
\newcommand{\ben}{\begin{enumerate}}
\newcommand{\een}{\end{enumerate}}

\newcommand{\blem}{\begin{lemma}}
\newcommand{\elem}{\end{lemma}}
\newcommand{\bthm}{\begin{theorem}}
\newcommand{\ethm}{\end{theorem}}
\newcommand{\bdefn}{\begin{definition}}
\newcommand{\edefn}{\end{definition}}
\newcommand{\bpf}{\begin{proof}}
\newcommand{\epf}{\end{proof}}
\newcommand{\bcor}{\begin{corollary}}
\newcommand{\ecor}{\end{corollary}}
\newcommand{\bprop}{\begin{proposition}}
\newcommand{\eprop}{\end{proposition}}

\newtheorem{proposition}{Proposition}
\newtheorem{lemma}{Lemma}
\newtheorem{theorem}{Theorem}

\newtheorem{remark}{Remark}
\newtheorem{definition}{Definition}
\newtheorem{corollary}{Corollary}

\makeatletter

\author{Adnan Raja,  and Pramod Viswanath}

\title{Compress-and-Forward Scheme for Relay Networks:  Backword Decoding and Connection to Bisubmodular Flows}

\begin{document}
\maketitle

\begin{abstract}

In this paper, a compress-and-forward scheme with backward decoding is presented for the unicast wireless relay network. 
The encoding at the source and relay is a generalization of the noisy network coding scheme (NNC) \cite{LKEC10}. While it achieves the same reliable data rate as noisy network coding scheme, the backward decoding allows for a better decoding complexity as compared to the joint decoding of the NNC scheme.
Characterizing the layered decoding scheme is shown to be equivalent to characterizing an information flow for the wireless network. 
A node-flow for a graph with bisubmodular capacity constraints is presented and a max-flow min-cut theorem is presented. This generalizes many well-known results of flows over capacity constrained graphs studied in computer science literature.
The results for the unicast relay network are generalized to the network with multiple sources with independent messages intended for a single destination.
\end{abstract}

\section{Introduction}

The primary focus of this paper is a unicast wireless relay network:  a single source node intends to communicate reliably with a single destination node with the assistance of many relay nodes. The communication channels are wireless; transmitted signals from a node are {\em broadcasted} to all other nodes;  received signals at a node is a linear {\em superposition} of the transmit signals with a random additive noise, which has the familiar Gaussian distribution.

In \cite{ADT11} a quantize-map-forward scheme was presented for the wireless relay network. It was shown that this scheme is approximately optimal, i.e.~it gives a reliability criterion for rates within a constant gap of the cutset bound, where the constant gap depends only on the size of the network and not on the channel parameters.  In this scheme, each node quantizes the received signal, symbol by symbol, at the noise level. The quantized symbols accumulated together in a block are then mapped to a transmit codeword at that node. These transmission codebooks at every node are generated independently of each other. 

In \cite{OD10}, a related scheme was presented for the wireless relay network.  Here, the coding and quantization is done in a structured manner using lattices. The scheme was shown to achieve performance similar to the quantize-map-forward scheme of \cite{ADT11} in terms of the reliable rates.

In \cite{LKEC10}, a {\em noisy network coding} scheme in the more general setting of the discrete memoryless network was presented for the unicast relay network and also generalized to the case of multicast and multiple sources with single destination.  
In this scheme,  the relay quantizes the received signal in blocks using vector-quantization, subsequently mapping each quantized codeword to a unique codeword, which is re-transmitted by the relay.
Specialized to the wireless network, the noisy network coding can be thought of as a vector version of the quantize-map-forward scheme, where each relay does a vector quantization rather than the scalar quantization proposed in \cite{ADT11}.

In \cite{AK11}, an alternate approach was provided, wherein the discrete superposition network was used as a digital interface for the wireless network and the scheme was constructed by lifting the scheme for the discrete superposition network. The discrete superposition network provided the quantization interface for this scheme. 

In this paper, a compress-and-forward scheme is presented for a relay network in the general setting of the discrete memoryless network. 
This encoding is similar to the noisy network scheme, but the relay mapping is generalized, so that the relay node compresses the received signal in blocks, on top of the vector quantization in NNC. 
The additional compression does not increase the achievable rate beyond the rate achievable by NNC; however, the first main result of this paper is that, if the compression rates are chosen appropriately then a lower complexity backward decoding achieves approximately the same rate. 
The above result was also proved independently in \cite{WX11}.
The second important result in this paper is to show that this appropriate choice of compression rates can be computed efficiently by computing a node-flow on a bisubmodular capacitated graph. The flow formulation captures the rate of {\em actual} information that should flow through each node to support a given rate of flow of information from the source to the destination. In other words, this paper shows that backward decoding does almost as good as joint decoding, if the relay nodes compress their signals to capture the right amount of information that should flow through that given the network topology.  

The paper presents a max-flow min-cut result for a node-flow on a bisubmodular capacitated graph. This is related to many well-known results of flows over capacity constrained graphs studied in computer science literature, albeit with two differences; the first one being that the flow is defined over nodes rather than the conventional approach of defining over edges; and the second is that the graphs are restricted to layered graphs alone. 
The first difference is a fundamental difference. Flows over graphs are conventionally defined as numbers over edges of the graph, such that for every node the incoming-flow is equal to the outgoing-flow. Since the motivation here is to model the wireless network where there are no physical edges, it is more appropriate to define node-flow rather than edge-flow; the relation being that the node-flow represents the incoming-flow or outgoing-flow at the node.  
The second is less fundamental and the restriction to layered graphs is done only because the block-coding scheme for the relay network can be studied by considering a virtual layered network. the layering offers a convenient way of defining the bisubmodular capacity functions on the layered graph.      

\begin{figure}
\scalebox{0.5}{\input{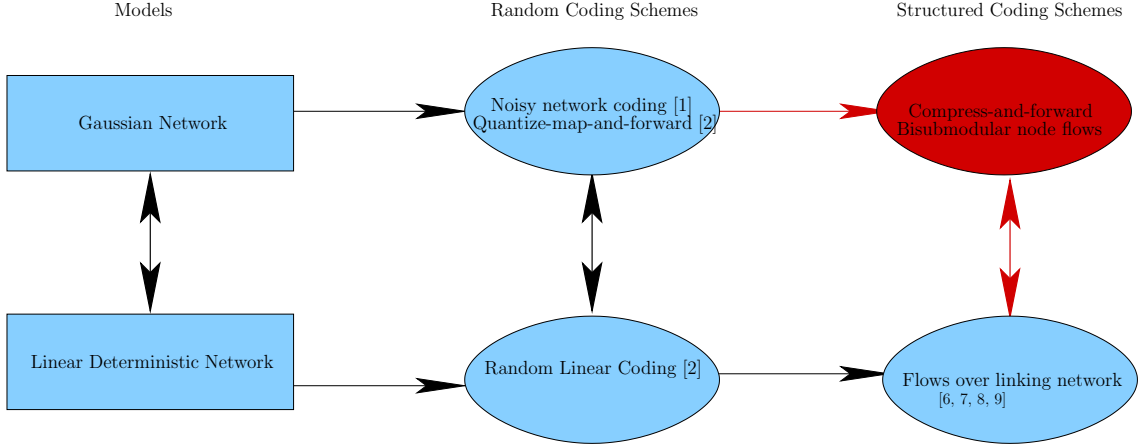}}
\caption{A depiction of the communication schemes on the Gaussian and linear deterministic networks. The main result of this paper is represented by the upper-right bubble in red.}
\label{fig:literature}
\end{figure}

The bisubmodular capacitated graph presented here is motivated by the ideas of linking systems and flows introduced in \cite{AF09, EF09, YS11, GIZ11} in the context of the linear deterministic network.
The linear deterministic network was introduced in \cite{ADT11} as a model that captures many features of the wireless network. Random coding argument was used to show the existence of schemes that achieve capacity of the linear deterministic network \cite{LKEC10, ADT11}.
On the other hand \cite{AF09, EF09} developed a polynomial time algorithm that discovers the relay encoding strategy using a notion of linear independence between channels.
Taking this concept forward, in \cite{YS11, GIZ11}, the concept of flow was introduced for the linear deterministic network. The flow value at each node in this network corresponds to the number of independent equations, that particular node needs to forward. 
The result in this paper can be viewed as a loose analog of these results in the context of the Gaussian network; see Figure~\ref{fig:literature}.
The additional structure of the linear deterministic channel, is used in \cite{AF09,EF09,YS11,GIZ11} to show that a single-block coding scheme where a simple permutation matrix at each node mapping the received vector to the transmit vector is optimal.  Both the flow values at the node and the permutation mapping were constructed in polynomial time. 

The rest of the paper is organized as follows.
In Section \ref{sec:comfwd} the compress-and-forward scheme for the relay network is described and characterized. A lower-complexity layered decoding is presented and the achievable rates are characterized. It is shown that this decoding scheme does as well as the joint decoding scheme. To prove this result, the notion of node-flows for a bisubmodular capacitated graph is developed in Section \ref{sec:flow}. 
In Section \ref{sec:multiSource}, the results are generalized to the network with multiple sources with independent messages intended for a single destination. In Section \ref{sec:spCases} we discuss the ramifications of our algebraic flow formulation to the important special cases of the Gaussian wireless relay network and the deterministic relay network.

\section{Unicast Relay Network}\label{sec:comfwd}

A communication network is represented by a set of nodes $\mc{V}$. Each node in the network abstracts a {\em radio}, which can both transmit and receive (in full or half duplex modes). The traffic is {\em unicast}: a single source node is
communicating reliably to a single destination node using the other nodes in the network as relays. We will be interested in a single-source single-destination relay network, which has a unique source node $s$ and destination node $d$ and the other nodes function as relay nodes.
At any node $v$, the transmit alphabet  is given by $\mc{X}_{v}$ and the receive alphabet by $\mc{Y}_{v}$ (supposed to be discrete sets, for the most part).
Time is discrete and synchronized among all nodes. The transmit symbol at any time at a node $v$ is given by $x_{v} \in \mc{X}_{v}$ and the receive symbol is given by $y_{v} \in \mc{Y}_{v}$.  {\em Memoryless} network will be considered here wherein the received symbol at any node at any given time depends (in a random fashion) only on the current transmitted symbols at other nodes.

A $(2^{TR},T)$ coding scheme for the relay
network, which communicates over $T$ time instants,
comprises of the following.  \ben
\item The {\em message} $W$, which is modeled as an independent random variable distributed uniformly on $[2^{TR}]$. $W$ is known at the source node and is intended for the destination node.
\item The {\em source mapping} for each time $t\in[T]$, \beqa f_{s,t}: (W \times \mc{Y}_{s}^{t-1}) \rightarrow \mc{X}_s.  \eeqa
\item The {\em relay mappings} for each $v \in \mc{V} \backslash \lbr s \rbr$ and $t \in [T]$, \beqa f_{v,t} : \mc{Y}_v^{t-1} \rightarrow \mc{X}_v. \label{eq:genRelMap} \eeqa
\item The {\em decoding map} at destination $d$, \beqa g_{d} : \mc{Y}_{d}^T \rightarrow \hat{W}. \eeqa
\een

The probability of error for destination $d$ under this coding
scheme is given by \beqa P_e  & \df & \Pr \{ \hat{W} \neq W
\}. \eeqa A rate $R$ (in bits per unit time)
is said to be achievable if for any $\epsilon >0 $, there exists a
$(2^{TR},T)$ scheme that achieves a
probability of error lesser than $\epsilon$ for all nodes, i.e.,
$P_e \leq \epsilon$. The capacity of the network is the supremum of all achievable rates.

It was shown in \cite{ADT11} that any arbitrary communication network can be converted into a layered network by coding over blocks of time. Each layer then captures the operations in the corresponding block of time. Further, if the nodes have half-duplex constraint, then this time-layering is done with a fixed transmit-receive schedule, which says which nodes are transmitting and which ones are listening in any block of time.
It is then a secondary question to optimize over the schedule in order to get the maximum rate of transmission.

\begin{figure}[htb]
\begin{center}
\scalebox{0.5}{\input{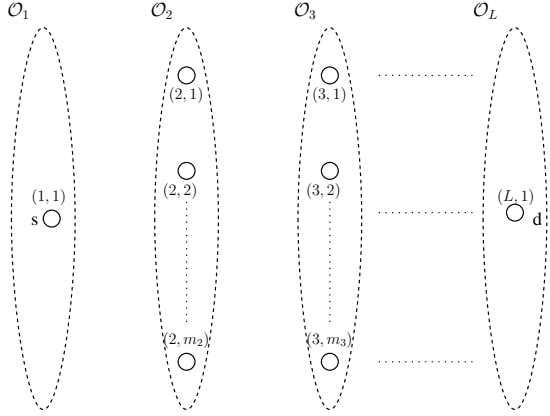}}
\end{center}
\caption{A layered network.} \label{fig:Lnetwork}
\end{figure}

Henceforth, the focus will be only on an $L$-layered network as shown in Figure \ref{fig:Lnetwork}, so that
\beq \mc{V} = \bigcup_{l=1}^{L}\mc{O}_{l}, \eeq
where $\mc{O}_{l}$ denotes the $m_{l}$ nodes in the $l$-th layer.
The $k$-th node in the $l$-th layer will be denoted by the ordered pair $(l,k)$.
The first layer has only one node which is the source node and is denoted by $(1,1)$ or $s$. The last layer has only the destination node and is denoted by $(L,1)$ or $d$. The nodes other than the source and the destination node will be referred to as the relay nodes and are denoted by $\mc{V}_{r}$, i.e.,
\beq \mc{V}_{r} = \bigcup_{l=2}^{L-1}\mc{O}_{l}. \eeq

In the layered network, the received symbol for a node in the $l+1$-th layer depends only on the transmit symbol from the nodes in the $l$-th layer. Therefore, for the layered network the channel which is denoted by a transition probability function can be simplified into a product across layers as follows:
\beq
p\lp y_{\mc{V}}|x_{\mc{V}}\rp = \prod_{l=1}^{L-1} p\lp y_{\mc{O}_{l+1}}|x_{\mc{O}_{l}} \rp.
\eeq
The noise across each relay node is assumed to be independent, which implies that the channel function for each layer is further given by, 
\beq \label{eq:chtrpr}
p(y_{\mc{O}_{l+1}}|x_{\mc{O}_{l}}) = \prod_{k=1}^{m_{l+1}} p(y_{(l+1,k)} | x_{\mc{O}_{l}}).
\eeq
Here $x_{\mc{O}_{l}}$ is used to denote $\lbr x_v :  v\in\mc{O}_{l} \rbr$. $y_{\mc{O}_{l}}$'s are similarly defined. This models the {\em communication channel} for the layered network.

In particular, if the received symbol is a deterministic function of the transmitted symbols, i.e.,
\beq y_{\mc{O}_{l+1}} =   g_{l}\lp x_{\mc{O}_{l}} \rp,  \label{eq:DetCh} \eeq
then the network is called a {\em deterministic network}. 
Further, if the transmit and received symbols are restricted to vectors over finite fields and the deterministic function is modeled as a linear function, such that 
\beq y_{\mc{O}_{l+1}} =   G_{l}x_{\mc{O}_{l}},  \eeq
then the network is called a {\em linear deterministic network}. If the network is a wireless network, then the alphabet sets are complex and the probability transition function linear with an additive complex Gaussian noise $z_{v}$, such that,
\beq y_{v} = \sum_{u\in\mc{O}_{l}} h_{v,u} x_{u} + z_{v}, \label{eq:WCh}\eeq
where $v\in\mc{O}_{l+1}$.
The wireless network is the one with the most practical interest and in \cite{ADT11} it was shown that the linear deterministic network captures many features of the wireless network.

\subsection{Compress-and-Forward Scheme} \label{sec:comFwdscheme}

In this section, the compress-and-forward scheme is described and it's performance is characterized.
It is a block-encoded scheme where each node performs its operation over blocks of time symbols.
The relay node quantizes (or compresses) the symbols it receives over a block of time to finite bits. These bits are then transmitted in the next block. The compression rate at a relay node is defined to be the rate of transmission of the compressed bits. 

Assuming that uniformly sized blocks of $T$ symbols are used by each node for this operation, a compress-and-forward scheme is 
parametrized by $\lp T, R, \lbr r_{v} \rbr_{v\in\mc{V}_{r}} \rp$, where $R$ is the overall rate of communication and $r_{v}$'s are the compression rates at the relay nodes. A rate vector  $\lp R, \lbr r_{v} \rbr_{v\in\mc{V}_{r}} \rp$ is said to be feasible w.r.t.~the compress-and-forward scheme, if for any arbitrary $\epsilon >0$, there exists a compress-and-forward scheme $\lp T, R, \lbr r_{v} \rbr_{v\in\mc{V}_{r}} \rp$ which achieves a probability of error less than $\epsilon$.
  
The following theorem characterizes the feasible region of $\lp R,\lbr r_{v} \rbr_{v\in\mc{V}_{r}} \rp$ for the compress-and-forward scheme.

\bthm \label{thm:cfRate}
A rate vector  $\lp R, \lbr r_{v} \rbr_{v\in\mc{V}_{r}} \rp$ is feasible if for some collection of random variables $\lbr X_{\mc{V}}, \hat{Y}_{\mc{V}} \rbr$, henceforth denoted by $Q_{p}$, which is distributed as
\beq 
p(X_{\mc{V}}, \hat{Y}_{\mc{V}},{Y}_{\mc{V}}) =  \lp \prod_{v\in\mc{V}} p(X_{v}) \rp p({Y}_{\mc{V}}|X_{\mc{V}}) \lp \prod_{v\in\mc{V}} p(\hat{Y}_{v} | Y_{v})\rp, \label{eq:RVcodebook}
\eeq
the vector $\lp R, \lbr r_{v} \rbr_{v\in\mc{V}_{r}} \rp$ satisfies
\beq
R <  r(\Omega^{c} \backslash \Phi) + I ( \hat{Y}_{\Phi} ; X_{\Omega} | X_{\Omega^{c}} ) - I ( \hat{Y}_{\Phi^{c} } ; Y_{\Phi^{c}} | X_{\mc{V}}),\label{eq:comfwd}
\eeq
$\forall~\Omega,\Phi,\;\textrm{s.t.,}\;S\in\Omega\subseteq \mc{V}, D\in\Phi \subseteq \Omega^{c}$,
where $r(\mc{A}) \df \sum_{v\in \mc{A}} r_{v}$.
\ethm

\noindent Note 1: The choice $\hat{Y}_{D} =Y_{D}$ is always optimal for \eqref{eq:comfwd}. 

\noindent Note 2: In the usual cut-set definition, the node-set is partitioned into two sets; a set containing the source $\Omega$ and the complementary set $\Omega^{c}$, containing the destination. However, here the node set is partition into a set containing the source - $\Omega$, a set containing the destination - $\Omega^{c}$, and the rest.

\bpf
The proof is by random coding technique. A random ensemble of coding scheme is defined using the collection of random variables $Q_{p}$ distributed as given by \eqref{eq:RVcodebook}.
A scheme in the ensemble is generated as follows. 

\ben
\item {\em Source codebook and encoding:} For each message $w\in[2^{TR}]$, the source generates a $T$-length sequence $x^{T}_{s}(w)$ using i.i.d.~$p(X_{S})$. 
\item {\em Relay codebooks and mappings:}
For every relay node $v \in \mc{V}_{r}$ a binned quantization codebook is generated with $2^{Tr_{v}}$ bins. The binned quantization codebook is given by $\hat{y}^{T}_{v}({w}_v,\bar{w}_v)$, where ${w}_{v} \in [2^{Tr_{v}}]$ and $  \bar{w}_{v} \in [ 2^{T\bar{r}_{v}}]$. And it is generated using i.i.d.~$p(\hat{Y}_{v})$.

Every relay node also generates a transmission codebook of size $2^{Tr_{v}}$, which consists of $x_{v}^{T}(w_{v})$ sequences generated using i.i.d.~$p(X_{v})$.

On receiving $y_{v}^{T}$, the relay node finds a vector $\hat{y}^{T}_{v}({w}_{v},\bar{w}_{v})$ in the quantization codebook that is jointly typical with $y_{v}^{T}$, and transmits $x_{v}^{T}(w_{v})$ corresponding to the bin number of the quantization vector.

If the relay cannot find any quantization vector, it transmits a sequence corresponding to any bin uniformly at random.
The probability that this latter event is arbitrarily is small is ensured by letting
\beq
\bar{r}_{v} = I(Y_{v},\hat{Y}_{v}) - r_{v} + \epsilon_{1}, \label{eq:Qrate}
\eeq
for an arbitrarily small $\epsilon_{1} > 0$.
This ensures that the total size of the quantization codebook is of the order $2^{T I(Y_{v},\hat{Y}_{v})}$. 

\item {\em Decoding:}
On receiving $y^{T}_{D}$, the destination node finds a {\em unique} $\hat{w}$, and {\em any} $\lbr (\hat{w}_{v},\hat{\bar{w}}_{v}) \rbr_{v\in\mc{V}_{r}}$, such that
\beq \lp x^{T}_{S}(\hat{w}), \lbr \hat{Y}^{T}_{v}(\hat{w}_{v},\hat{\bar{w}}_{v}), {x}^{T}_{v}(\hat{w}_{v}) \rbr_{v\in\mc{V}_{r}}, y^{T}_{D} \rp \in \mc{T}_{\epsilon}^{T}.
\label{eq:decode_condition}
\eeq
If it is successful, the destination declares $\hat{w}$ as the decoded message; if not, the destination declares an error. 

\een

The theorem follows by the standard argument of showing that the average probability of error, averaged over the ensemble of codes and over all messages, goes to $0$ as $T$ tends to infinity.  
The details of the error probability analysis are in Appendix \ref{app:cfRate}.

\epf

In the usual communication problem setup, one is interested in only maximizing the overall communication rate $R$. 
The following corollary of the above theorem establishes the achievable rate by the compress-and-forward scheme.
\bcor
The communication rate $R$ is achievable by the compress-and-forward scheme if
\beq
R < \min_{\Omega \subseteq \mc{V},S\in\Omega} I ( \hat{Y}_{\Omega^{c}} ; X_{\Omega} | X_{\Omega^{c}} ) - I ( \hat{Y}_{\Omega } ; Y_{\Omega} | X_{\mc{V}},), \label{eq:R-comfwd}
\eeq
for some collection of random variables $Q_{p}$.
\ecor

\bpf
The compress-and-forward scheme with $R_{v} = I(Y_{v},\hat{Y}_{v}) + \epsilon_{1}$ achieves this rate.
\epf

It should be noted that the achievable rate in \eqref{eq:R-comfwd} is the same as the one obtained in noisy network coding scheme in \cite{LKEC10}. This is not surprising as by allowing the compression rates to be large enough, the scheme essentially reduces to the noisy network coding scheme, where every quantized codeword is uniquely mapped to a re-transmission codeword at the relay node.

\subsection{A low-complexity layered decoding scheme}

A maximum likelihood decoder maximizes the probability of the received vector conditioned on the transmitted codeword at the source. 
(Note that the jointly-typical-set decoding is a proof technique for the random coding argument and it upper-bounds the error probability that can be achieved by the maximum likelihood (ML) decoder.
\beq
\textrm{ML decoder:}\quad \hat{w} = \textrm{argmax}_{w} p\lp y_{D}^{T}|x_{S}^{T}(w)\rp. 
\eeq 
The conditional probability depends on the channel model and the operations (quantization, compression and mapping) at each node. Therefore implementing a ML decoder has very high complexity. In \cite{NWJTN10}, the ML decoder is implemented for a simple one-relay network with binary LDPC codes and a reduced quantizer operation for which the decoding reduces to belief-propagation over a large Tanner graph, which comprises the Tanner graphs of the LDPC codes for each node, the quantization and mapping operation, and the network itself. Even when this simplified encoding scheme is extended to a network with multiple layers of relay nodes, the decoding complexity would be large. 
In this section, a simplified decoding architecture is presented for the compress-and-forward scheme which operates layer-by-layer and decodes the compressed bits transmitted by each relay node. 

{\em Layered decoding scheme:} The decoder at the destination node operates backwards layer-by-layer. 
First, it decodes the messages (or compressed bits) transmitted by the nodes in the layer $\mc{O}_{L-1}$. Then using these decoded messages, it decodes the messages in the layer $\mc{O}_{L-2}$. This process continues till the destination node eventually decodes the source message. 
Note that the layered decoding scheme is the same as the backward decoding for the block-encoding schemes in relay networks. 

The following theorem characterizes the feasible region of $\lp R,\lbr r_{v} \rbr_{v\in\mc{V}_{r}} \rp$.

\bthm \label{thm:cfRateRDec}
A rate vector  $\lp R, \lbr r_{v} \rbr_{v\in\mc{V}_{r}} \rp$ is feasible for the compress-and-forward scheme, under the layered decoding scheme, if for some $Q_{p}$ the vector $\lp R, \lbr r_{v} \rbr_{v\in\mc{V}_{r}} \rp$ satisfies 
\begin{align}
r(U) &\leq I(X_{U};Y_{D}|X_{\mc{O}_{L-1}\backslash U}),\;\forall\;U\subseteq \mc{O}_{L-1},\label{eq:cfRateRDec1}\\
r(U) - r(\mc{O}_{l+1}\backslash V) &\leq I(X_{U};\hat{Y}_{V} | X_{\mc{O}_{l}\backslash U}) - I(\hat{Y}_{\mc{O}_{l+1}\backslash V};Y_{\mc{O}_{l+1}\backslash V}| X_{\mc{O}_{l}}),\nn\\
&\qquad\qquad\qquad\forall\;U\subseteq \mc{O}_{l},V\subseteq \mc{O}_{l+1},2\leq l\leq L-2 ,\label{eq:cfRateRDec2}\\
R - r(\mc{O}_{2}\backslash V) &\leq I(X_{S};\hat{Y}_{V}) - I(\hat{Y}_{\mc{O}_{l+1}\backslash V};Y_{\mc{O}_{l+1}\backslash V}| X_{S}),\;\forall V\subseteq \mc{O}_{2}.\label{eq:cfRateRDec3}
\end{align}
\ethm

\bpf

\begin{figure}[htb]
\begin{center}
\scalebox{0.5}{\input{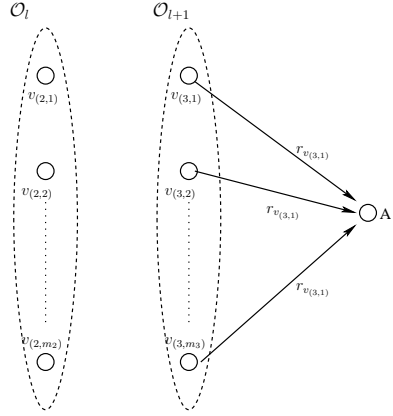}}
\end{center}
\caption{A hypothetical network.} \label{fig:localNetwork}
\end{figure}

The proof is by backward induction. Assuming that the destination has decoded the messages transmitted by the relay nodes in layer $\mc{O}_{l+1}$, the probability of error for decoding the messages from the layer $\mc{O}_{l}$ is considered. To do so, a hypothetical layered network as shown in Figure~\ref{fig:localNetwork} is considered. This network consists of the layers $\mc{O}_{l}$ and $\mc{O}_{l+1}$ and in addition a layer with an aggregator node $A$. A node $v_{(l+1,j)}$ in layer $\mc{O}_{l+1}$ is connected to the aggregator node with wired link of capacity $r_{v_{(l+1,j)}}$ bits per symbol. This layer represents the forward part of the network beyond layer $\mc{O}_{l+1}$. 

This network is now a multiple-source single-destination relay network, with all the nodes in layer $O_{l}$ being source nodes and the aggregator node as the destination node. The node $v_{(l,j)}$ has a message for the aggregator node with rate $r_{v_{(l,j)}}$. The noisy network coding scheme \cite{LKEC10} assures that the messages can be decoded with arbitrarily small probability of error, if
\beq
r(U) - r(\mc{O}_{l+1}\backslash V) \leq I(X_{U};\hat{Y}_{V} | X_{\mc{O}_{l}\backslash U}) - I(\hat{Y}_{V^{c}};Y_{V^{c}}| X_{\mc{O}_{l}}),
\eeq
$\forall\;U\subseteq \mc{O}_{l},V\subseteq \mc{O}_{l+1}$, where the above inequality corresponds to the cut $\Omega = U \bigcup V^{c}$.

\epf
 
Note that the layered decoding scheme is weaker than the ML decoding scheme. Therefore the feasible region under the layered decoding scheme should be a strict subset of the feasible region under the ML decoding scheme. 

However, the following theorem shows that the compress-and-forward scheme with layered decoding achieves similar communication rate as the noisy network coding scheme. 

\bthm \label{thm:cfEffRateRDec}
The communication rate $R$ is achievable by the compress-and-forward scheme with layered decoding if for some collection of random variables $Q_{p}$,
\beq
R < \min_{\Omega \subseteq \mc{V},S\in\Omega} I ( \hat{Y}_{\Omega^{c}} ; X_{\Omega} | X_{\Omega^{c}} ) - \kappa_{1}, \label{eq:R-comfwd-layered}
\eeq
where the constant $\kappa_{1}$ is given by the recursive relation,  
\beq
\kappa_{l} =  I(\hat{Y}_{\mc{O}_{l+1}};Y_{\mc{O}_{l+1}}| X_{\mc{O}_{l}})+\kappa_{l+1}|\mc{O}_{l+1}|, \label{eq:kappa-cf}
\eeq
and $\kappa_{L-1}=0$.
\ethm
\bpf
The above theorem will be proved by characterizing an {\em information flow} for the network in the Section \ref{sec:cfEffRateRDec}.
\epf

Note that the conditions of Theorem \ref{thm:cfRateRDec} can be interpreted as a flow decomposition for the layered network. If $R$ is the information that flows from the source to the destination, then the flow decomposition gives the effective amount of information that flows through each node. 
If the compression rate at each relay node is made approximately equal to the information flowing through that node, then the layered decoding where the destination ends up decoding the effective information at each node has a chance to work. Thus, in order to choose the right compression rates at each node, a flow decomposition for the network must be obtained. These notions are made more precise in the next section. 

\begin{remark}
Assuming the maximum likelihood (ML) decoding is done by an exhaustive search as given by \eqref{eq:decode_condition}, the decoding complexity of the joint decoding is the product of the codebooks of all the nodes. Therefore the complexity of the joint-decoding is given by
\beq
\mathfrak{C}_{\textrm{joint}} = 2^{RT}\prod_{v\in\mc{V}_{r}} n_{Q,v},
\eeq
where $n_{Q,v}$ is the number of quantization points in the relay quantization codebook. 
With the compress-and-forward scheme with the layered decoding, the complexity is reduced to 
\beq
\mathfrak{C}_{\textrm{layered}} = \sum_{l=1}^{L-1}2^{r(\mc{O}_{l})T}\prod_{v\in\mc{O}_{l+1}} n_{Q,v}.
\eeq
\end{remark}

\section{Flows with Bisubmodular Capacity Constraints} \label{sec:flow}

Maximum flow problems are extensively studied in graph theory and combinatorial optimization \cite{S03}. The problems are most often motivated from the study of transportation and communication networks. 
A directed graph $(\mc{V},\mc{E})$ consists of the set of vertices or nodes $\mc{V}$ and the set of edges $\mc{E} \subseteq \mc{V}\times \mc{V}$. 
Traditionally, flow is defined to be a non-negative function over the set of all edges which satisfy the {\em flow-conservation law} at each vertex other than the source and the destination node. Further, the flow over any edge is less than the capacity of that the edge.   
The classic max-flow min-cut result of \cite{FF56} characterizes the maximum flow from the source to destination node and shows it to be equal to the min-cut of the graph. 
In order to distinguish from the concept of the node-flow that will be introduced here, such a flow is called an edge flow over an edge-capacitated graph. Beginning from the single commodity result of \cite{FF56}, various extensions of these problems have been considered. In particular, the edge-capacitated graph was extended to a polymatroidal network \cite{LM82}, where the flow is constrained not only by the edge-capacities but by joint capacities on sets of incoming and outgoing edges at every vertex. A special case is the node-capacitated graph\cite{FHL08}, where the constraints on the flow are on the sum-total of the incoming and outgoing flow at each node.

In this section, the concept of a node-flow in the context of a layered graph with bisubmodular constraints on the flows is introduced. The node-flows can be related to the edge-flows with flow-conservation at the node. Note that the conservation law for edge-flow enforces that the net incoming flow at any node is equal to the net outgoing flow at the node and this quantity can be viewed as the node-flow for a node.  The bisubmodular constraints can be viewed as generalizations of the polymatroidal constraints of \cite{LM82}. 
The definitions here are motivated by the layered coding scheme for the wireless network, which was presented in the previous chapter.  
The main result is a max-flow min-cut theorem for the single-commodity node-flow for a graph with bisubmodular capacity constraints.  
The result is closely related to, and can be viewed as a generalization of, the flow introduced in the context of the linear deterministic networks and polylinking systems in \cite{YS11,GIZ11}. 

\subsection{A max-flow min-cut theorem}

In this section, the max-flow min-cut theorem is proved for {\em single-commodity node-flow} on a {\em layered graph} with {\em bisubmodular capacity} constraints. 
 
{\em Layered graph:} A layered graph is considered, which is represented by a set of nodes $\mc{V}$, which can be decomposed into subsets $\mc{O}_{l}, 1\leq l\leq L$ as shown in Figure \ref{fig:Lnetwork}. 
The layering is ensured by the edges of the graph, which connect nodes in any layer $l$ to nodes in the subsequent layer $l+1$. Since the edges do not play any role in the problem here, beyond ensuring the layering, they will henceforth be neglected. 
The first layer $\mc{O}_{1}$ has a single node, which is the source node and the last layer $\mc{O}_{L}$ has a single node, which is the destination node.

{\em Bisubmodular capacity functions:} The bisubmodular capacity functions are defined for the layered graph using a family of $L-1$ functions  \newline $\lbr \rho_{l} : 1\leq l \leq L-1 \rbr$,  $\rho_{l}:2^{\mc{O}_{l}}\times 2^{\mc{O}_{l+1}} \rightarrow \real^{+}$,  which satisfy the following properties:
\ben
\item $\rho_{l}$ is bisubmodular, i.e., $\forall U_1,U_2\subseteq \mc{O}_l, V_1,V_2\subseteq  \mc{O}_{l+1}$, 
\beq
\rho_{l}(U_{1} \cup U_{2}, V_{1} \cap V_{2}) + \rho_{l}(U_{1} \cap U_{2}, V_{1} \cup V_{2}) \leq \rho_{l}(U_{1}, V_{1}) + \rho_{l}(U_{2}, V_{2}).
\eeq
\item $\rho_{l}$  is non-decreasing, i.e.
\beq \rho_l(U,V) \leq \rho_l (U_{1},V_{1}),~\text{for}~ U \cup V \subseteq U_{1} \cup V_{1}.\eeq
\item If $U = \emptyset$ or $V=\emptyset$, then \beq \rho_{l}(U,V) = 0. \eeq
\een

{\em Node-flow:}  The node-flow for the layered graph is defined as a function $f:\mc{V}\rightarrow \real^{+}$ which satisfies the capacity constraints, i.e.,
\beq f(V) - f(\mc{O}_{l}\backslash U)\leq \rho_{l}(U,V),\qquad\forall~U\subseteq \mc{O}_{l},\;V\subseteq \mc{O}_{l+1},\forall l\in[L-1],\eeq
where $f(A)$ is an over-loaded notation, such that when $A\subseteq \mc{V}$ then $ f(A) \df \sum_{v\in A} f(v)$.
Further, the destination node must sink the flow from the source. Therefore $f(D) = f(S)$.  

The max-flow problem is to find the maximum $f(S)$ that can be supported given the capacity constraints on the graph. An efficient algorithm to compute the flow at each node given any $f(S)$ that can be supported is also sought.    

An upper bound on the max-flow is given by the cut function. 
 
{\em Cut function}: The cut function ${C}:2^{\mc{V}}\rightarrow\real_{+}$ is defined as
\beq C(\Omega) \df \sum_{l=1}^{L-1} \rho_{l}(\Omega_{l},\mc{O}_{l+1}\backslash \Omega_{l+1}), \eeq
where $\Omega_{l} \df \Omega \cap \mc{O}_{l}$.

Clearly, 
\beq 
\max f(S) \leq \min_{\Omega\subseteq\mc{V}} {C}(\Omega).
\eeq 

The next theorem shows that the min-cut is achievable. The proof is constructive and gives and efficient method of computing the flow. 
\bthm \label{thm:flow}
\beq\max f(S) = \min_{\Omega\subseteq\mc{V}} {C}(\Omega).\eeq
\ethm
 
\bpf
The proof is based on the polymatroid intersection theorem. The details are in Appendix \ref{apx:proofFlow}. 
\epf

The max-flow min-cut theorem for node-flows with bisubmodular constraints presented here is closely related to the max-flow min-cut results of \cite{YS11, GIZ11}. 
\cite{YS11} considered linear deterministic networks, which led to bisubmodular capacity functions arising from the rank of a matrix. \cite{GIZ11} considered polylinking systems, where the bisubmodular capacity functions are given by the polylinking function. The results of \cite{GIZ11} generalized the results of \cite{YS11} by showing that a linear deterministic network is a special case of polylinking system. 

The max-flow min-cut theorem can be easily generalized to the following two cases:
\beit \item {\em Multi-source:} Consider a layered graph with $J$ source nodes in $\mc{O}_{1}$ and a single destination node in $\mc{O}_{L}$, such that $f(\mc{O}_{1}) = f(D)$. For this case, the following corollary generalizes Theorem \ref{thm:flow}.
\bcor
$\lbr f(v) | v\in \mc{O}_{1} \rbr$ is a feasible flow iff,
\beq f(\Omega_{1}) \leq {C}(\Omega),\qquad\forall\;\Omega\subseteq\mc{V},\eeq  
where $\Omega_{1}\df\Omega \cap \mc{O}_{1}$.  
\ecor 
\item {\em Multi-destination:} Consider a layered graph with a single source node in $\mc{O}_{1}$ and J destination nodes in $\mc{O}_{L}$, such that $f(S) = f(\mc{O}_{L})$. For this case, the following corollary generalizes Theorem \ref{thm:flow}.
\bcor
$\lbr f(v) | v\in \mc{O}_{L} \rbr$ is a feasible flow iff,
\beq f(\Omega_{L}) \leq {C}(\Omega),\qquad\forall\;\Omega\subseteq\mc{V},\eeq  
where $\Omega_{L}\df\Omega \cap \mc{O}_{L}$.  
\ecor
\eeit

Note that the proof for the multiple sources (or destinations) case follows by adding a hypothetical supernode $A$ in layer $0$ (or $L+1$) with capacity functions $\rho_{0}$ (or $\rho_{L}$) given by $\rho_{0}(A,V) = \sum f(v),\;\forall\;V\subseteq\mc{O}_{1}$ (or  $\rho_{L}(V,A) = \sum f(v),\;\forall\;V\subseteq\mc{O}_{L}$). 

\subsection{Proof of Theorem \ref{thm:cfEffRateRDec}: A Compress-and-Forward Scheme from Flows} \label{sec:cfEffRateRDec}

In this section, Theorem \ref{thm:cfEffRateRDec} is proved by establishing a connection between the compression rates of the compress-and-forward scheme with the layered decoding and the node-flows with bisubmodularity constraints. Recall that the achievable rates for the compress-and-forward with the layered decoding scheme are given by \eqref{eq:cfRateRDec1}-\eqref{eq:cfRateRDec3}, which appear very much like the bisubmodular capacity constraints. 

To make this connection more precise, first observe the following proposition. 

 \begin{proposition}
 \label{prop:mi_channelfunction}
 Given the collection of random variables $Q_{p}$ distributed as given by \eqref{eq:RVcodebook}, the family of $L-1$ functions $\rho_{l}:\mc{O}_{l}\times\mc{O}_{l+1}\rightarrow\real^{+},\;\forall l\in[L-1]$ defined by \beq  \rho_{l}(U,V) \df I(X_{U};\hat{Y}_{V} | X_{\mc{O}_{l}\backslash U}) \label{eq:MI-bisubmodcapfn} \eeq
forms a family of bisubmodular capacity functions.
 \end{proposition}
\bpf
Appendix \ref{app:mi_channelfunction}.
\epf
For any $\Omega\subseteq\mc{V}$, the corresponding cut value $C(\Omega)$ is now given by
\beqa
C(\Omega) &=& \sum_{l=1}^{L-1} I(X_{\Omega_{l}};\hat{Y}_{\mc{O}_{l+1}\backslash \Omega_{l+1}}| X_{\mc{O}_{l}\backslash \Omega_{l}})  \\
& = & I(\hat{Y}_{\Omega_{c}};X_{\Omega} | X_{\Omega^{c}}).
\eeqa
Theorem \ref{thm:flow} is then used construct a flow $f(v)$ for this network, such that 
\beq f(S)  \leq \min_{\Omega} I(\hat{Y}_{\Omega_{c}};X_{\Omega} | X_{\Omega^{c}}),\qquad S\in\Omega, D\in\Omega^{c},  \eeq
and 
\beq f(V) - f(\mc{O}_{l}\backslash U)\leq \rho_{l}(U,V),\qquad\forall~U\subseteq \mc{O}_{l},\;V\subseteq \mc{O}_{l+1},\forall l\in[L-1]. \eeq

For any $v\in\mc{O}_{l},l\in[L-1]$, let 
 \beq 
 r_{v} = f(v) - \kappa_{l}, 
 \eeq 
 and $R = f(S) - \kappa_{1}$, 
where $\kappa_{l}$ is given by \eqref{eq:kappa-cf}.

Then  $\forall U\neq\emptyset\subseteq\mc{O}_{l},\;V\subseteq\mc{O}_{l+1}$,
\begin{align}
r(U) - r(\mc{O}_{l+1}\backslash V) &= f(U) - f(\mc{O}_{l+1}\backslash V) - |U|\kappa_{l}+|\mc{O}_{l+1}\backslash V|\kappa_{l+1} \\
&\leq \rho_{l}(U,V) - \kappa_{l}+|\mc{O}_{l+1}|\kappa_{l+1} \\
&= \rho_{l}(U,V) - I(\hat{Y}_{\mc{O}_{l+1}}; {Y}_{\mc{O}_{l+1}} | X_{\mc{O}_{l}}) \\
&\leq I(X_{U};\hat{Y}_{V} | X_{\mc{O}_{l}\backslash U}) - I(\hat{Y}_{\mc{O}_{l+1}\backslash V}; {Y}_{\mc{O}_{l+1}\backslash V} | X_{\mc{O}_{l}}).
\end{align} 

Therefore $\lp R, \lbr r_{v} \rbr_{v\in\mc{V}_{r}} \rp$ satisfies \eqref{eq:cfRateRDec1}-\eqref{eq:cfRateRDec3}. 
This proves Theorem \ref{thm:cfEffRateRDec}.

\section{Generalizations to multi-source networks} \label{sec:multiSource}

\begin{figure}[htb]
\begin{center}
\scalebox{0.7}{\input{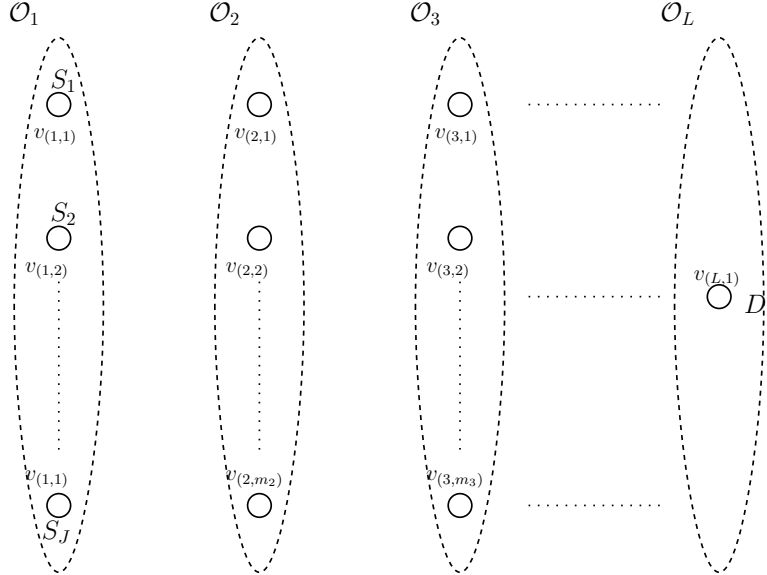}}
\end{center}
\caption{A layered multi-source network.} \label{fig:Lnetwork-MC}
\end{figure}

The communication network with multiple source nodes $\lbr S_{i} | i\in [J] \rbr$ is illustrated in Figure \ref{fig:Lnetwork-MC}. The source node $S_{i}$ has independent message $W_{i}$ at rate $R_{i}$. There is a common destination node $D$. 
The multi-source relay network was perhaps first studied in \cite{PerronThesis09, PDTita09}, where the rate region for the deterministic case and an
approximate rate region for the Gaussian case were established.
The noisy network coding scheme of \cite{LKEC10} extends to this case as well. In fact this result was used for each layer to analyze the layered decoding scheme in the proof of Theorem \ref{thm:cfRateRDec}. 

The results of the compress-and-forward scheme and the layered decoding scheme can be generalized to the communication network with multiple source nodes and a common destination node.

The following corollary extends the results of the compress-and-forward scheme for the unicast network to the multi-source relay network. 

\bthm \label{thm:cfRate-MS}

The communication rates $\vec{R} = \lp R_{1},\ldots,R_{J} \rp$ are achievable by the compress-and-forward scheme (with joint decoding) for the multi-source single destination network if, for some collection of random variables $Q_{p}$ which is distributed as \eqref{eq:RVcodebook}, the rates satisfy
\beq
R(\Omega_{1}) <  I ( \hat{Y}_{\Omega^{c}} ; X_{\Omega} | X_{\Omega^{c}} ) - I ( \hat{Y}_{\Omega } ; Y_{\Omega} | X_{\mc{V}},), \quad\forall~\Omega,\;\textrm{s.t.,}\;\Omega\subseteq \mc{V}, D\in\Omega^{c}, \label{eq:R-comfwd-MS}
\eeq
where $\Omega_{1}\df\Omega\cap\mc{O}_{1}$.

Further, with the layered decoding scheme, the rates $\vec{R} = \lp R_{1},\ldots,R_{J} \rp$  are achievable if
\begin{align}
R(\Omega_{1}) <  I ( \hat{Y}_{\Omega^{c}} ; X_{\Omega} | X_{\Omega^{c}} ) - |\Omega_{1}| \kappa_{1}, 
\end{align}
where $\kappa_{1}$ is given by \eqref{eq:kappa-cf}.
\ethm
 
The results can be proved by adding a hypothetical supernode in layer $0$, which is connected to the source nodes with orthogonal wired links such that the wired link to node $S_{i}$ is of rate $R_{i}$.  

\section{Special cases} \label{sec:spCases}

\subsection{Wireless network}

For the special case of the Wireless network described by \eqref{eq:WCh}, the achievable rates can be compared to the cutset bound \cite{elemIT}. 

As noted in \cite{LKEC10}, a good choice for $\hat{Y}_{v}$ for the Gaussian network is given by 
\beq \hat{Y}_{v} = Y_{v} + \hat{Z}_{v},  \eeq
where $\hat{Z}_{v}\sim\mc{CN}(0,1)$ is independent across nodes.

The particular choice of $\hat{Y}_{v}$ implies that the quantization is done at the noise level. This also agrees with the philosophy in \cite{ADT11, AK11}, where the quantization was done at the noise level to show approximate optimality;  in \cite{ADT11}, scalar quantization was done at the noise level, and in \cite{AK11}, quantization was done using the discrete superposition network, which was a model obtained from the wireless network by clipping the signal at the noise level. 

As shown in \cite{LKEC10}, with this choice of $\hat{Y}_{v}$ and with $X_{\mc{V}}\sim\mc{CN}({0,I})$,
\begin{align} 
I ( \hat{Y}_{\Omega^{c}} ; X_{\Omega} | X_{\Omega^{c}} )  &= \log \left | I+\frac{H_{\Omega\Omega^{c}}H_{\Omega\Omega^{c}}^{*}}{2} \right | \\
& \geq \log \left | I+H_{\Omega\Omega^{c}}H_{\Omega\Omega^{c}}^{*} \right |  - \frac{| \Omega^{c} |}{2}. \label{eq:inBoundEq1}
\end{align}
And further,
\beq I ( \hat{Y}_{v} ; Y_{v} | X_{\mc{V}} ) \leq 1.\label{eq:inBoundEq2}\eeq

Using \eqref{eq:inBoundEq1}, \eqref{eq:inBoundEq2} and Lemma 6.6 in \cite{ADT11}, the following corollary of Theorem \ref{thm:cfRate-MS}  follows.

\bcor
If  $\vec{R} = \lp R_{1},\ldots,R_{J} \rp$ is in the cutset bound, then rates $\vec{R} - 3|\mc{V}|\vec{1}$ are achievable by the compress-and-forward scheme (with joint decoding) for the multi-source single destination Gaussian network. 
Further, with the layered decoding scheme, the rates $\vec{R} - (2|\mc{V}|+\kappa_{1}^{g})\vec{1}$  are achievable, where 
\beq
\kappa_{l}^{g} =  1+\kappa_{l+1}^{g}|\mc{O}_{l+1}|,
\eeq
and $\kappa_{L-1}^{g}=0$. 
\ecor

\subsection{Deterministic network}

For the special case of the deterministic network described by \eqref{eq:DetCh}, the optimal choice of $\hat{Y}_{v}$ is $Y_{v}$ and with this choice 
\beq I ( \hat{Y}_{\Omega^{c}} ; X_{\Omega} | X_{\Omega^{c}} ) = H ( \hat{Y}_{\Omega^{c}} | X_{\Omega^{c}} ). \label{eq:inBoundEq1-det}\eeq
And further,
\beq I ( \hat{Y}_{v} ; Y_{v} | X_{\mc{V}} ) = 0.\label{eq:inBoundEq2-det}\eeq

Therefore, specializing the results of Theorem \ref{thm:cfRate-MS} leads to the following corollary.
\bcor
For the multi-source single-destination deterministic network, 
$\vec{R} = \lp R_{1},\ldots,R_{J} \rp$ is achievable by the compress-and-forward scheme with the layered decoding scheme if for some collection of random variables $Q_{p}$ which is distributed as \eqref{eq:RVcodebook},
\beq
\vec{R} \in \bar{\mc{C}}(Q_{p}),\label{eq:R-comfwd-MS-det}
\eeq
where $\bar{\mc{C}}(Q_{p})$ is the cutset bound evaluated under the product distribution for the network \cite{ADT11}. 
\ecor 

Specializing further to the linear deterministic region, it can be shown that the product distribution (with uniformly distributed $X_{v}$ over all input alphabets) maximizes the cutset bound, thereby showing that all rates in the cutset bound are achievable.

\section{Conclusion}

In this paper, the compress-and-forward scheme is analyzed for the unicast relay network. It is shown that while it achieves the same overall rate as NNC, it allows for a lower complexity layered/backward decoding algorithm. However, this requires each relay node to compress their information to the right amount. This paper also presents a computationally efficient way of finding the optimal compression rates at each relay node using a node-flow formulation over a bisubmodular constrained graph.

\appendix

\section{Probability of Error Analysis for CF Scheme} \label{app:cfRate}

Without loss of generality we assume that the message with index $1$ is transmitted at the source and the index corresponding to the quantized vectors at each node is $(1,1)$.
We will find the probability of error that this message is wrongly decoded at the destination.
We denote by $\mc{E}_{w, \lp w,\bar{w}\rp_{\mc{V}_{r}} }$ the event that
\beq \lp x^{T}_{s}({w}), \lbr \hat{y}^{T}_{(l,k)}({w}_{(l,k)},{\bar{w}}_{(l,k)}), {x}^{T}_{(l,k)}({w}_{(l,k)}) \rbr_{(l,k)\in\mc{V}_{r}}, y^{T}_{d} \rp \in \mc{T}_{\epsilon}^{T}. \eeq
Here $\lp w,\bar{w}\rp_{\mc{V}_{r}}$ is shorthand for $\lbr \lp w_{v},\bar{w}_{v}\rp | v\in\mc{V}_{r} \rbr$. 
The error event is the union of two terms and is given by 
\beq
\lp \bigcup_{w_{\mc{V}_{r}},\bar{w}_{\mc{V}_{r}}} \mc{E}_{1,(w,\bar{w})_{\mc{V}_{r}}} \rp^{c} \bigcup \lp \bigcup_{w\neq 1,w_{\mc{V}_{r}},\bar{w}_{\mc{V}_{r}}} \mc{E}_{w,(w,\bar{w})_{\mc{V}_{r}}} \rp.
\eeq
The first term corresponds to the event that the transmitted message is not jointly typical and the second term corresponds to some other message other than the transmitted being jointly typical. 
The first event can be upper bounded by $\mc{E}_{1,(1,1)_{\mc{V}_{r}}}^{c}$.
For any $\Omega \subseteq \mc{V}_{r}$, and $\Phi \subseteq  \mc{V}_{r}\backslash\Omega$, let 
\begin{align}
\mathfrak{S}_{\Omega,\Phi} \df \lbr (w,(w,\bar{w})_{\mc{V}_{r}}) |  w \neq 1, \right. & w_{(l,k)}\neq 1 \forall(l,k) \in \Omega, \nn \\
& w_{(l,k)}=1,\bar{w}_{(l,k)} \neq 1 \forall (l,k) \in \Omega^{c}\backslash\Phi, \nn \\
& \left. w_{(l,k)}=1,\bar{w}_{(l,k)} = 1 \forall (l,k) \in \Phi \rbr,
\end{align}
and
\beq
\mc{E}_{\Omega,\Phi} \df \bigcup_{\mathfrak{S}_{\Omega,\Phi}} \mc{E}_{w,(w,\bar{w})_{\mc{V}_{r}}}.
\eeq
The second event can be equivalently written as, 
\beq 
\lp \bigcup_{w\neq 1,w_{\mc{V}_{r}},\bar{w}_{\mc{V}_{r}}} \mc{E}_{w,(w,\bar{w})_{\mc{V}_{r}}} \rp = \bigcup_{\Omega,\Phi} \mc{E}_{\Omega,\Phi},
\eeq
The probability or error by union bound can be upper bounded by,
\beq
\prob(error) \leq \prob(\mc{E}_{1,(1,1)_{\mc{V}_{r}}}^{c}) + \sum_{\Omega,\Phi} \prob\lp \mc{E}_{\Omega,\Phi} \rp.
\eeq
From the properties of joint typicality, it can be shown that the first term goes to $0$ and $T\rightarrow\infty$. 
It can be shown that
\beqan
\prob\lp \mc{E}_{\Omega,\Phi} \rp & \eqo & 2^{T \lp R + r(\Omega) + \bar{r}(\Phi^{c}) \rp} 2^{T \lp  H(Y_{d}, \hat{Y}_{\Phi}, \hat{Y}_{\Phi^{c}}, X_{\Omega}, X_{\Omega^{c}}, X_{s}) - H(X_{\Omega}, X_{s}) - H(Y_{d}, \hat{Y}_{\Phi},X_{\Omega^{c}}) - \sum_{(l,k)\in\Phi^{c}}  H(\hat{Y}_{(l,k)}) \rp } \\
&=& 2^{T \lp R + r(\Omega) + \bar{r}(\Phi^{c}) \rp} 2^{T \lp  H(Y_{d}, \hat{Y}_{\Phi}, \hat{Y}_{\Phi^{c}} | X_{\Omega}, X_{\Omega^{c}}, X_{s})  - H(Y_{d}, \hat{Y}_{\Phi} | X_{\Omega^{c}}) - \sum_{(l,k)\in\Phi^{c}}  H(\hat{Y}_{(l,k)}) \rp }\\
&=& 2^{T \lp R + r(\Omega) + \bar{r}(\Phi^{c}) \rp} 2^{-T \lp   H(Y_{d}, \hat{Y}_{\Phi} | X_{\Omega^{c}}) - H(Y_{d}, \hat{Y}_{\Phi} | X_{\Omega}, X_{\Omega^{c}}, X_{s}) + \sum_{(l,k)\in\Phi^{c}}  H(\hat{Y}_{(,k)}) - H(\hat{Y}_{\Phi^{c}} | X_{\Omega}, X_{\Omega^{c}}, X_{s}) \rp } \\
& = & 2^{T \lp R + r(\Omega) + \bar{r}(\Phi^{c}) \rp} 2^{-T \lp  I(Y_{d}, \hat{Y}_{\Phi}; X_{\Omega}, X_{s} | X_{\Omega^{c}}) + \sum_{(l,k)\in\Phi^{c}} I(\hat{Y}_{(l,k)} ; X_{\mc{V}_{r}},X_{s}) \rp }.
\eeqan
Here $r(A) \df \sum_{v\in A}r_{v}$. 
Using the Markovian property of the random variables, we have that
\beq
I(\hat{Y}_{(l,k)} ; X_{\mc{V}_{r}},X_{s}) = I(\hat{Y}_{(l,k)} ; {Y}_{(l,k)}) - I(\hat{Y}_{(l,k)} ; {Y}_{(l,k)} | X_{\mc{V}_{r}},X_{s} ),
\eeq
and using \eqref{eq:Qrate} we have
\beq
\prob\lp \mc{E}_{\Omega,\Phi} \rp = 2^{T \lp R - r(\Omega^{c} \backslash \Phi) - I(Y_{d}, \hat{Y}_{\Phi}; X_{\Omega}, X_{s} | X_{\Omega^{c}}) + I(\hat{Y}_{\Phi^{c}} ; {Y}_{\Phi^{c}} | X_{\mc{V}_{r}},X_{s} ) \rp}.
\eeq
Therefore $\prob\lp \mc{E}_{\Omega,\Phi} \rp \rightarrow 0$, if
\beq
R < r(\Omega^{c} \backslash \Phi) + I(Y_{d}, \hat{Y}_{\Phi}; X_{\Omega}, X_{s} | X_{\Omega^{c}}) - I(\hat{Y}_{\Phi^{c}} ; {Y}_{\Phi^{c}} | X_{\mc{V}_{r}},X_{s} ).
\eeq

\section{Proof of Theorem \ref{thm:flow}}\label{apx:proofFlow}

The theorem will be proved in a slightly general setting, allowing multiple nodes in layer $\mc{O}_{1}$ and layer $\mc{O}_{L}$.  
Assuming that the flow values for these layers $\mc{O}_{1}$ and $\mc{O}_{L}$ are given and satisfy 
\begin{align}
f(\mc{O}_{1}) &= f(\mc{O}_{L}), \\
f(\Omega_{1}) - f(\Omega_{L}) &\leq  C(\Omega),\quad\forall\;\Omega \subseteq \mc{V}, \label{eq:boundaryflowconstraint}
\end{align}
the flow for all intermediate layers will be constructed. 

The proof is by inductive construction. 

For L=2, there are no intermediate layers and the theorem holds by definition.
Consider $L>2$. The induction hypothesis assumes that the flow can be constructed with fewer than $L$ layers and the flow for the boundary layers are specified with the constraints given by \eqref{eq:boundaryflowconstraint}

Consider any $L_{0} \in \lbr 2,\ldots,L-1 \rbr$.
Define networks $\mc{N}_{A}$ and $\mc{N}_{B}$ to be the sub-networks of $\mc{N}$ with the set of vertices $\mc{V}_{A} = \cup_{l=1}^{L_{0}} \mc{O}_{l}$ and $\mc{V}_{B} = \cup_{l=L_{0}}^{L} \mc{O}_{l}$ respectively. Similarly, denote the cut for the two networks by $C_{A}$ and $C_{B}$ respectively.

Next, a flow for the layer $\mc{O}_{L_{0}}$ will be constructed which satisfies the following conditions. 
\beqa
f(\mc{O}_{L_{0}}) &=& f(\mc{O}_{1}), \label{eq:sumflow} \\
f(\Omega_{A} \cap \mc{O}_{1}) - f(\Omega_{A} \cap \mc{O}_{L_{0}}) &\leq& C_{A}(\Omega_{A}),~\forall~\Omega_{A} \subseteq \mc{V}_{A}, \text{ and} \label{eq:flowA}\\
f(\Omega_{B} \cap \mc{O}_{L_{0}}) - f(\Omega_{B} \cap \mc{O}_{L}) &\leq& C_{B}(\Omega_{B}),~\forall~\Omega_{B} \subseteq \mc{V}_{B}. \label{eq:flowB}
\eeqa
The induction hypothesis would then guarantee that the flows for the intermediate layers in the sub-networks $\mc{N}_{A}$ and $\mc{N}_{B}$ can be constructed.

Using \eqref{eq:sumflow}, the set of linear inequalities given by \eqref{eq:flowA} can be written as, 
\beqa
f(\Omega_{A}^{c} \cap \mc{O}_{L_{0}}) - f(\Omega_{A}^{c} \cap \mc{O}_{1}) &\leq& C_{A}(\Omega_{A}),~\forall~\Omega_{A} \subseteq \mc{V}_{A}, 
\eeqa
where $\Omega_{A}^{c}=\mc{V}_{A}\backslash\Omega_{A}$. For any fixed $T\subseteq\mc{O}_{L_{0}}$, the collection of inequalities where $\Omega_{A}^{c} \cap \mc{O}_{L_{0}} = T$, can be concisely represented as, 
\beqa
f(T) &\leq&  \min \lbr C_{A}(\Omega_{A}) + f(\Omega_{A}^{c} \cap \mc{O}_{1}) : \Omega_{A}^{c} \cap \mc{O}_{L_{0}} = T\rbr.
\eeqa
Defining
\beq
r_{A}(T) \df \min \lbr C_{A}(\Omega_{A}) + f(\Omega_{A}^{c} \cap \mc{O}_{1}) : \Omega_{A}^{c} \cap \mc{O}_{L_{0}} = T\rbr,
\eeq
the set of linear inequalities given by \eqref{eq:flowA} can be concisely written as, 
\beq
f(T) \leq r_{A}(T),\qquad\forall~T\subseteq\mc{O}_{L_{0}}.
\eeq
Similary, defining
\beq
r_{B}(T) \df \min \lbr C_{B}(\Omega_{B}) + f(\Omega_{B} \cap \mc{O}_{L}) : \Omega_{B} \cap \mc{O}_{L_{0}} = T\rbr,
\eeq
the set of linear inequalities given by \eqref{eq:flowB} can be concisely written as, 
\beq
f(T) \leq r_{B}(T),\qquad\forall~T\subseteq\mc{O}_{L_{0}}.
\eeq

The following properties for the functions $r_{A}(T)$ and $r_{B}(T)$ can be established.
\blem \label{lem:polyf}
The functions $r_{A}(T)$ and $r_{B}(T)$ are \begin{itemize}
 \item submodular,
 \item  non-decreasing, and
 \item satisfy  $r_{A}(\emptyset) = 0$ and $r_{B}(\emptyset) = 0$.
 \end{itemize}
\elem
\bpf Appendix \ref{app:polyf}. \epf 
Define the following polymatroids with the functions $r_{A}$ and $r_{B}$.
\beqa
P_{A} &=& \lbr \mbf{x} \in \real^{m_{L_{0}}}_{+} : x(U) \leq r_{A}(U),~\forall~U\in\mc{O}_{L_{0}} \rbr \\
P_{B} &=& \lbr \mbf{x} \in \real^{m_{L_{0}}}_{+} : x(U) \leq r_{B}(U),~\forall~U\in\mc{O}_{L_{0}} \rbr,
\eeqa
where $\mbf{x} = [x(1) \ldots x(m_{L_{0}})]$ and $x(U) \df \sum_{u\in U} x(u)$.
The conditions \eqref{eq:sumflow}-\eqref{eq:flowB} are now equivalent to finding
\beqa
[f(L_{0},1) \ldots f(L_{0},m_{L_{0}})] \in P_{A} \cap P_{B},
\eeqa
such that $f(\mc{O}_{L_{0}}) = f(\mc{O}_{1})$.
It then follows from Edmond's polymatroid intersection (\cite{S03}, Corollary 46.1c) that:
\beq \text{max} \lbr x(\mc{O}_{L_{0}}) : \mbf{x} \in P_{A} \cap P_{B} \rbr  = \min_{T\subseteq\mc{O}_{L_{0}}} \lbr r_{A}(\mc{O}_{L_{0}} \backslash T) + r_{B}(T) \rbr. \label{eq:polyXthm} \eeq
Therefore the required flow exists since
\beqa
f(\mc{O}_{1}) &\leq& \min_{T\subseteq\mc{O}_{L_{0}}} \lbr r_{A}(\mc{O}_{L_{0}} \backslash T) + r_{B}(T) \rbr \\
&=& \min_{\Omega\in\mc{V}} \lbr C(\Omega) + f(\mc{O}_{1}\backslash \Omega_{1}) + f(\Omega_{L}) \rbr.
\eeqa
Further, in Theorem 47.1 of \cite{S03} it is shown that the maximizing $\mbf{x}$ in \eqref{eq:polyXthm} can be computed in {\em polynomial} time in the dimension of $\mbf{x}$. Hence, the flow can also be computed in polynomial time in the number of nodes.

\section{ Proof of Lemma~\ref{lem:polyf} \label{app:polyf}}

We will prove the lemma for $r_{B}(T)$. The proof for $r_{A}(T)$ is similar.

\ben
\item Submodularity:

Let,
\beqa
r_{B}(T^{(1)}) &=& C_{B}(\Omega_{B}^{(1)}) + d(\Omega^{(1)}_{B} \cap \mc{O}_{L}), \quad \Omega_{B}^{(1)} \cap \mc{O}_{L_{0}} = T^{(1)} \label{eq:lem1}\\
r_{B}(T^{(2)}) &=& C_{B}(\Omega_{B}^{(2)}) + d(\Omega^{(2)}_{B} \cap \mc{O}_{L}), \quad \Omega_{B}^{(2)} \cap \mc{O}_{L_{0}} = T^{(1)}. \label{eq:lem2}
\eeqa
Since,
\beqa
(\Omega_{B}^{(1)} \cup \Omega_{B}^{(2)})  \cap \mc{O}_{L_{0}} &=& T^{(1)} \cup T^{(2)}, \\
(\Omega_{B}^{(1)} \cap \Omega_{B}^{(2)})  \cap \mc{O}_{L_{0}} &=& T^{(1)} \cap T^{(2)},
\eeqa
it follows that
\beqa
r_{B}(T^{(1)} \cup T^{(2)}) &\leq& C_{B}(\Omega_{B}^{(1)} \cup \Omega_{B}^{(2)}) + d( (\Omega^{(1)}_{B} \cup \Omega_{B}^{(2)})  \cap \mc{O}_{L}), \label{eq:lem3} \\
r_{B}(T^{(1)} \cap T^{(2)}) &\leq& C_{B}(\Omega_{B}^{(1)} \cap \Omega_{B}^{(2)}) + d( (\Omega^{(1)}_{B} \cap \Omega_{B}^{(2)})  \cap \mc{O}_{L}). \label{eq:lem4}
\eeqa
By definition of cut and the bi-submodularity of $\rho_{l}$, it is easy to verify that $C_{B}(\Omega_{B})$ is submodular.
And since $d$ is an additive function, it then follows that $r_{B}(T)$ is sub modular.

\item Non-decreasing:

Consider $T^{(1)} \subseteq T^{(2)}$. Let
\beqa
r_{B}(T^{(1)}) &=& C_{B}(\Omega_{B}^{(1)}) + d(\Omega^{(1)}_{B} \cap \mc{O}_{L}), \quad \Omega_{B}^{(1)} \cap \mc{O}_{L_{0}} = T^{(1)}.
\eeqa
Let $\Omega_{B} = \Omega_{B}^{(1)} \cup T^{(2)}\backslash T^{(1)} \supseteq \Omega_{B}^{(1)} $, so that $\Omega_{B} \cap \mc{O}_{L_{0}} = T^{(2)}$.
By the definition of cut and the non-decreasing property of $\rho_{l}$, it follows that
$ C_{B}(\Omega_{B}^{(1)}) \leq C_{B}(\Omega_{B})$. Also $d(\Omega^{(1)}_{B} \cap \mc{O}_{L})\leq d(\Omega_{B}\cap \mc{O}_{L})$.
Therefore
\beqa
r_{B}(T^{(2)}) &=& C_{B}(\Omega_{B}) + d(\Omega_{B} \cap \mc{O}_{L}) \\
&\geq& C_{B}(\Omega_{B}^{(1)}) + d(\Omega^{(1)}_{B} \cap \mc{O}_{L}) \\
&=& r_{B}(T^{(1)}).
\eeqa

\item $r_{B}(\emptyset) = 0$:

When $T=\emptyset$, by letting $\Omega_{B}=\emptyset$, it follows that $r_{B}(\emptyset)=0$.

\een

\section{Proof of Proposition \ref{prop:mi_channelfunction}} \label{app:mi_channelfunction}

We need to show that $I(X_{U};\hat{Y}_{W} | X_{\mc{O}_{l}\backslash U})$ satisfies the three properties of channel functions. Firstly we show that it is bi-submodular.

\beqa
I(X_{U};\hat{Y}_{W} | X_{\mc{O}_{l}\backslash U}) &=& H(\hat{Y}_{W} | X_{\mc{O}_{l}\backslash U}) - H(\hat{Y}_{W} | X_{\mc{O}_{l}}) \\
& =& H(\hat{Y}_{W}, X_{\mc{O}_{l}\backslash U}) - H(X_{\mc{O}_{l}\backslash U}) - H(\hat{Y}_{W} | X_{\mc{O}_{l}}).
\eeqa
The submodularity of entropy \cite{KK80} implies that $H(\hat{Y}_{W}, X_{\mc{O}_{l}\backslash U})$ is bi-submodular.

The submodularity of entropy follows from the fact that given
collection of random variables $\Upsilon_{1}$ and $\Upsilon_{2}$, we have
\beqa
H(\Upsilon_{1}) + H(\Upsilon_{2}) - H(\Upsilon_{1} \cup \Upsilon_{2}) - H(\Upsilon_{1} \cap \Upsilon_{2})  &=& I(\Upsilon_{1} \backslash \Upsilon_{2} ; \Upsilon_{2} \backslash \Upsilon_{1} | \Upsilon_{1} \cap \Upsilon_{2}) \\
&\geq& 0.
\eeqa
The product form of the random variables implies that  $H(X_{\mc{O}_{l}\backslash U})$ and $H(\hat{Y}_{W} | X_{\mc{O}_{l}})$ are modular or additive.
Therefore, $I(X_{U};\hat{Y}_{W} | X_{\mc{O}_{l}\backslash U})$ is bi-submodular.

Next, we show the non-decreasing property. Given $U_{1} \subseteq U \subseteq \mc{O}_{l}$ and $W_{1} \subseteq W\subseteq \mc{O}_{l+1}$, we have
\beqa
I(X_{U};\hat{Y}_{W} | X_{\mc{O}_{l}\backslash U}) &=& H(X_{U}|X_{\mc{O}_{l}\backslash U}) - H(X_{U}|X_{\mc{O}_{l}\backslash U}\hat{Y}_{W}) \\
&\geq& H(X_{U}|X_{\mc{O}_{l}\backslash U}) - H(X_{U}|X_{\mc{O}_{l}\backslash U}\hat{Y}_{W_{1}}) \\
&=& I(X_{U};\hat{Y}_{W_{1}} | X_{\mc{O}_{l}\backslash U}) \\
&=& H(\hat{Y}_{W_{1}} | X_{\mc{O}_{l}\backslash U}) - H(\hat{Y}_{W_{1}} | X_{\mc{O}_{l}}) \\
&\geq& H(\hat{Y}_{W_{1}} | X_{\mc{O}_{l}\backslash U_{1}}) - H(\hat{Y}_{W_{1}} | X_{\mc{O}_{l}}) \\
& = & I(X_{U_{1}};\hat{Y}_{W_{1}} | X_{\mc{O}_{l}\backslash U_{1}}),
\eeqa
where both the inequalities follow from the fact that conditioning reduces entropy.

The third property is readily seen.

\section*{Acknowledgments}

The authors would like to thank Chandra Chekuri for the many useful discussions.

\bibliographystyle{IEEE_ECE}
\bibliography{references}

\end{document}